\documentclass[aps,pre,twocolumn,superscriptaddress,floatfix]{revtex4}
\usepackage{graphicx}
\usepackage[dvips,unicode,colorlinks,linkcolor=blue,citecolor=blue,urlcolor=blue]{hyperref}
\usepackage{amsmath}
\usepackage{amssymb}
\usepackage{natbib,hyperref}

\begin{document}

\preprint{APS/123-QED}

\title{Discrete Breathers and Multi-Breathers in Finite Vibro-Impact Chain}

\author{Itay Grinberg}

\email{GrinbergItay@gmail.com}

\author{Oleg V. Gendelman}

\email{ovgend@technion.ac.il}

\affiliation{Faculty of Mechanical Engineering\\
Technion - Israel Institute of Technology }

\date{\today}
\begin{abstract}
We explore dynamics of discrete breathers and multi-breathers in finite one-dimensional
chain. The model involves parabolic on-site potential with rigid constraints
and linear nearest-neighbor coupling. The rigid non-ideal impact
constraints are the only source of nonlinearity and damping in the
model. The model allows derivation of exact analytic solutions for
the breathers and multi-breathers with arbitrary set of localization sites,
both in conservative and forced-damped settings. We choose periodic boundary conditions;
exact solutions for other types of
the boundary conditions are also possible. Local character of the
nonlinearity allows explicit derivation of a monodromy matrix for
the breather solutions. Consequently, a stability of the derived
breather and multi-breather solutions can be efficiently studied in
the framework of simple methods of linear algebra, and with rather moderate computational efforts. We demonstrate
that finitness of the chain fragment and proximity of the localization
sites strongly effect existence and stability patterns of these localized
solutions.
\begin{description}
\item [{PACS~numbers}] 05.45.Yv, 63.20.Pw, 63.20.Ry{\small \par}
\end{description}
\end{abstract}



\maketitle

\section{Introduction}

Localization is a well\textendash known phenomenon in nonlinear lattices
\citep{Flach,Flach2,localization,localization2,localization3,localization4,localization5,localization6}.
Remarkable example of such localized dynamic states are discrete breathers (DBs),
sometimes referred to as intrinsic localized modes (ILM) or discrete
solitons. In lattices with linear coupling, the DBs are localized exponentially;
if the coupling is strongly nonlinear, the localization may become
hyper-exponential \citep{Flach2}. The DBs were experimentally observed
and explored in many physical systems, including, among
others, superconducting Josephson junctions\citep{Josephson}, nonlinear
magnetic metamaterials\citep{Metamaterials}, electrical lattices\citep{Electrical},
micro-mechanical cantilever arrays\citep{beam,beam2,beam3,beam4,beam5},
Bose-Einstein condensates\citep{bose}, and chains of mechanical oscillators\citep{chain,pendula,chain2}.

Theoretical investigation of the DBs relied primarily on numeric and
approximate analytic methods\citep{Flach,Flach2}. Exact analytic
solutions for the DBs are quite scarce, due to combination of discreteness
and nonlinearity. Known exceptions are completely integrable Ablowitz-Ladik
model\citep{AL} and chain with homogeneous interactions\citep{FO}.
Recently this family has been extended by derivation of analytic solutions for the
DBs in conservative vibro-impact chains \citep{chain}. The vibro-impact chains
also allowed computation of the exact solutions for
forced-damped discrete breathers \citep{chain2,PG}.

In these latter works, one important peculiarity of the vibro-impact
models has been explored and used. If all interactions besides the
collisions with the impact constraints are considered to be linear,
then it is possible to drastically simplify the analysis of stability for the DB solutions.
The DBs are periodic solutions of a system of ordinary differential equations, and their stability is determined
by the location of eigenvalues of the monodromy matrix, computed for
the DB solution in the given system\citep{floquet}. Commonly the
monodromy matrix has to be computed numerically, by  integration
of the system with $N$ degrees of freedom over the period, with $2N$
different initial conditions for every point in the space of parameters.
For systems large enough, such task is amenable only by supercomputers.
For the vibro-impact models mentioned above, the monodromy matrix
can be explicitly expressed in a general form\citep{chain2,PG}, and
the numeric part is reduced to relatively rapid and simple computation of the eigenvalues
of this matrix for given parameter values. This simplification allows
detailed exploration of the stability patterns in the space of parameters for the
DB solutions.

Current work is based on the approaches developed in \citep{chain,chain2,PG},
and extends them in two aspects. First of all, all experimental setups
mentioned above include finite (and sometimes
rather small) number of coupled oscillatory systems. From the other
side, it is possible to excite more than one site of the lattice,
and these excited sites are not necessarily adjacent. Thus, one can
observe and explore the multi-breather (MB) solutions. The existing knowledge on properties and especially on the stability of the MBs is rather limited.
Due to its simplicity, the vibro-impact model seems  natural for derivation and exploration of the MB solutions in finite systems. The paper addresses exactly this problem.
 In Section \ref{sec: II} we describe the general model settings. In Section \ref{sec: III} the exact
solutions for the multi-breathers both in Hamiltonian and forced-damped
settings are derived. Section \ref{sec: IV} investigates the stability
properties of the obtained solutions. Section \ref{sec:Numeric-Validation}
presents numeric validation and illustrations of the results of the previous Sections.
Section \ref{sec:VI} adds some concluding remarks.

\section{\label{sec: II}Description of the model}

We consider a finite chain of identical unit masses, coupled with
linear springs, and with periodic boundary conditions. Besides, each
mass has the same on-site interaction -- a linear spring with symmetric pair
of impact barriers located at distances $u_{n}=\pm1$ from the trivial equilibrium position. This unit scaling does not
restrict the generality. The Hamiltonian of the systems that includes $\left(N+1\right)$ masses is written
as follows:
\begin{equation}
\begin{array}{c}
H=\underset{n=0}{\overset{N}{\sum}}{\left(\cfrac{1}{2}p_{n}^{2}+V{\left(u_{n}\right)}\right)}+\\
+\underset{n=0}{\overset{N-1}{\sum}}{W{\left(u_{n}-u_{n+1}\right)}}+W{\left(u_{N}-u_{0}\right)}
\end{array}
\end{equation}
\begin{eqnarray}
V{\left(x\right)} & = & \begin{cases}
\cfrac{\gamma_{1}}{2}x^{2} & \,\,\,\,\,\left|x\right|<1\\
\mbox{infinity} & \,\,\,\,\,\left|x\right|=1
\end{cases}\\
W{\left(x\right)} & = & \cfrac{\gamma_{2}}{2}x^{2}
\end{eqnarray}
where $p_{n}=\dot{u}_{n}$ is the momentum of each particle, $\gamma_{1}$
and $\gamma_{2}$ are the on-site and coupling stiffnesses respectively
and $V{\left(x\right)}$ and $W{\left(x\right)}$ are the on-site
and coupling potentials respectively.

This yields the following linear equations of motion between the impacts, i.e. for $\left|u_{n}\right|<1$ for all particles:
\begin{eqnarray}
\ddot{u}_{0}+\gamma_{1}u_{0}+\gamma_{2}\left(2u_{0}-u_{1}-u_{N}\right) & = & 0\label{eq:4}\\
\ddot{u}_{n}+\gamma_{1}u_{n}+\gamma_{2}\left(2u_{k}-u_{k+1}-u_{k-1}\right) & = & 0\\
\ddot{u}_{N}+\gamma_{1}u_{N}+\gamma_{2}\left(2u_{N}-u_{0}-u_{N-1}\right) & = & 0
\end{eqnarray}

We adopt here traditional Newtonian model of the inelastic impacts.
Namely, when at certain time instance $t=t_{b}$ some particle
achieves the impact barrier ($u_{n}{\left(t_{b}\right)}=\pm1$), its
velocity is instantaneously modified according to the following law:

\begin{equation}
\dot{u}_{n}{\left(t_{b}+\right)}=-e\dot{u}_{n}{\left(t_{b}-\right)}\label{eq:7}
\end{equation}
Here $0<e\leq1$ is a restitution coefficient.

\section{\label{sec: III}Exact Solutions for the Multi-breathers}

\subsection{Hamiltonian Model}

Let us proceed with analytic solution for the multi-breathers in the vibro-impact
chain fragment described in the previous section . First, we are going
to consider the conservative case, where $e=1$ and no external force is applied. In the most generic
setting, the multi-breather solution in this system corresponds to
periodic oscillatory state, in which certain subset of masses periodically
impacts the barriers, and the others do not achieve them. Without
loss of generality, we suggest that the particle with $n=0$ is engaged
in the impacts and the particle with $n=N$ does not impact the constraints.
Every single impact may be presented as a result of action of the external
force in the form of delta-function. We also suggest that all impacting
masses undergo the impacts simultaneously. It is
possible that some masses impact their right barriers, and the others
impact their left barriers at the same time instance. Taking into
account the periodicity of the MB solution, the latter
should obey the following system of equations:

\begin{eqnarray}
\begin{array}{c}
\ddot{u}_{0}+\gamma_{1}u_{0}+\gamma_{2}\left(2u_{0}-u_{1}-u_{N}\right)=\\
=2p_{0}\delta_{0k}\underset{j=-\infty}{\overset{\infty}{\sum}}{\left(\delta{\left(t-\frac{\pi\left(2j+1\right)}{\omega}\right)}-\delta{\left(t-\frac{2\pi j}{\omega}\right)}\right)}
\end{array}\label{eq:8}\\
\begin{array}{c}
\ddot{u}_{n}+\gamma_{1}u_{n}+\gamma_{2}\left(2u_{k}-u_{k+1}-u_{k-1}\right)=\\
=2p_{k}\delta_{nk}\underset{j=-\infty}{\overset{\infty}{\sum}}{\left(\delta{\left(t-\frac{\pi\left(2j+1\right)}{\omega}\right)}-\delta{\left(t-\frac{2\pi j}{\omega}\right)}\right)}
\end{array}\\
\ddot{u}_{N}+\gamma_{1}u_{N}+\gamma_{2}\left(2u_{N}-u_{0}-u_{N-1}\right)=0\label{eq:10}
\end{eqnarray}
where $\delta_{nk}$ is  Kronecker delta, $\delta{\left(t\right)}$
is Dirak delta function, $k\in \left\{ 0,m_{1},m_{2},\cdots,m_{L}\right\} $,
$m_{l}$ are the indices of the impacting masses, $L+1$ is a number
of impacting particles, and $2p_{k}$ is an amount of momentum transferred
to the $k$-th particle by the impact constraint at the instance of the
impact. $\omega$ is a fundamental frequency of the breather ($T={2\pi}/{\omega}$ is a minimal period). For the conservative model, $\left|p_{k}\right|$ is also
the magnitude of the velocity of the impacting mass before and after
the impact, due to the unit restitution coefficient.

The periodicity of the impacts allows us to rewrite equations \eqref{eq:8}-\eqref{eq:10}
in terms of generalized Fourier series:
\begin{eqnarray}
\begin{array}{c}
\ddot{u}_{0}+\gamma_{1}u_{0}+\gamma_{2}\left(2u_{0}-u_{1}-u_{N}\right)=\\
=-\frac{4\omega p_{0}\delta_{0k}}{\pi}\underset{j=0}{\overset{\infty}{\sum}}{\cos{\left(\left(2j+1\right)\omega t\right)}}
\end{array}\label{eq:11}\\
\begin{array}{c}
\ddot{u}_{n}+\gamma_{1}u_{n}+\gamma_{2}\left(2u_{n}-u_{n+1}-u_{n-1}\right)=\\
=-\frac{4\omega p_{k}\delta_{nk}}{\pi}\underset{j=0}{\overset{\infty}{\sum}}{\cos{\left(\left(2j+1\right)\omega t\right)}}
\end{array}\label{eq:12}\\
\ddot{u}_{N}+\gamma_{1}u_{N}+\gamma_{2}\left(2u_{N}-u_{0}-u_{N-1}\right)=0\label{eq:13}
\end{eqnarray}

Heterogeneous solutions of equations \eqref{eq:11}-\eqref{eq:13} is presented
in the following form of Fourier series:
\begin{equation}
u_{n}=\sum_{j=0}^{\infty}{u_{n,j}\cos{\left(\left(2j+1\right)\omega t\right)}}\label{eq:14}
\end{equation}

Furthermore, since the solution should be localized, the following
anzats for $u_{n,j}$ is used:
\begin{equation}
u_{n,j}=\sum_{k}{\left(A_{j,k}f_{j}^{\left|n-k\right|}+B_{j,k}f_{j}^{-\left|n-k\right|}\right)}\label{eq:15}
\end{equation}

Physically, this form of solution corresponds to the exponential localization
around each breather site. System \eqref{eq:11}-\eqref{eq:13} between
the impact time instances is linear and solutions for coefficients
$f_{i}$ should obey the following relationships:
\begin{widetext}
\begin{equation}
\begin{array}{c}
f_{j}=\cfrac{\gamma_{1}+2\gamma_{2}-\left(2j+1\right)^{2}\omega^{2}\pm\sqrt{\left(\left(2j+1\right)^{2}\omega^{2}-\gamma_{1}-2\gamma_{2}\right)^{2}-4\gamma_{2}^{2}}}{2\gamma_{2}}=\\
=\cfrac{\gamma_{1}+2\gamma_{2}-\left(2j+1\right)^{2}\omega^{2}\pm\sqrt{\left(\left(2j+1\right)^{2}\omega^{2}-\gamma_{1}-4\gamma_{2}\right)\left(\left(2j+1\right)^{2}\omega^{2}-\gamma_{1}\right)}}{2\gamma_{2}}
\end{array}\label{eq:16}
\end{equation}

\end{widetext}

In order to make spatial localization possible, the term under the
square root should be positive -- in other terms, the Hamiltonian
DB exists only in the attenuation zone of the chain. As for the choice of the $\pm$
sign - it is easy to see that inversion of the sign does not modify the
solution. Relation between the coefficients $A_{j,k}$ and $B_{j,k}$
can be obtained from the periodic boundary conditions by substituting
\eqref{eq:15} in \eqref{eq:13}:
\begin{equation}
A_{j,k}=B_{j,k}f_{j}^{-N-1}
\end{equation}

It is important to note here that $A_{j,k}=0$ as $N\to\infty$ and
the solution converges to that of the infinite chain \citep{chain}.

Finally, for the impacting masses one obtains :
\begin{equation}
B_{j,k}=\frac{4\omega p_{k}}{\pi\gamma_{2}\left(f_{j}-f_{j}^{-1}\right)\left(f_{j}^{-N-1}-1\right)}
\end{equation}

Thus, the solution for the chain fragment with the MB is expressed in the following form:
\begin{equation}
u_{n}=\sum_{j=0}^{\infty}{u_{n,j}\cos{\left(\left(2j+1\right)\omega t\right)}}
\end{equation}
where
\begin{equation}
u_{n,j}=\sum_{k}{\frac{4\omega p_{k}\left(f_{j}^{\left|n-k\right|-N-1}+f_{j}^{-\left|n-k\right|}\right)}{\pi\gamma_{2}\left(f_{j}-f_{j}^{-1}\right)\left(f_{j}^{-N-1}-1\right)}}
\end{equation}

The only remaining unknown is the amount of momentum transferred in the course of each impact, i.e. $p_{k}$.
It can be computed, if one takes into account the location of the
barriers. Let $n=k_{s}$ be some impacting particle; then, at the
instance of the impact one obtains:
\begin{equation}
u_{k_{s}}{\left(0\right)}=\sum_{k}{\frac{4\omega p_{k}}{\pi\gamma_{2}}\chi_{k_{s},k}}=\pm1\label{eq:21}
\end{equation}
where the $\pm$ sign determines whether the specific mass is in-phase
or out-of-phase with respect to the other impacting masses and,
\begin{equation}
\chi_{k_{s},k}=\sum_{j=0}^{\infty}{\cfrac{\left(f_{j}^{\left|k_{s}-k\right|-N-1}+f_{j}^{-\left|k_{s}-k\right|}\right)}{\left(f_{j}-f_{j}^{-1}\right)\left(f_{j}^{-N-1}-1\right)}}
\end{equation}

Note that if the location of the impacting mass $k_{s}$ at $t=0$
is $-1$, one should obtain $p_{k_{s}}<0$ and vice versa.

The obtained set of equations can also be written in the following
more compact form:
\begin{equation}
\left[\begin{array}{c}
p_{0}\\
p_{k_{1}}\\
\vdots\\
p_{k_{m}}
\end{array}\right]=\frac{\pi\gamma_{2}}{4\omega}\mbox{C}^{-1}\left[\begin{array}{c}
1\\
1\\
\vdots\\
1
\end{array}\right]\label{eq:23}
\end{equation}
where $\left(m+1\right)\geq N$ is the number of impacting masses
and:
\begin{equation}
\mbox{C}=\left[\begin{array}{cccc}
\chi_{0,0} & \chi_{0,k_{1}} & \cdots & \chi_{0,k_{M}}\\
\chi_{k_{1},k_{1}} & \chi_{k_{1},k_{1}} &  & \vdots\\
\vdots &  & \ddots & \vdots\\
\chi_{k_{M},0} & \cdots & \cdots & \chi_{k_{M},k_{M}}
\end{array}\right]
\end{equation}

It is important to mention here that the self-consistency of the obtained solution (i.e. the
fact that the particles, which are not selected as the ``impacting''
ones, indeed do not achieve the constraints) is difficult to prove analytically
due to the complexity of the equations; it is explored numerically
in section \ref{sec:Numeric-Validation}. However, it is easy to see
that as $N\to\infty$ there is exponential localization with respect
to the localization sites, if they are concentrated at some finite
sub-fragment of the chain.

\subsection{Forced-Damped Model}

In this case the model is slightly altered, since all masses are subjected
to the same external force $F{\left(t\right)}$. We consider symmetric periodic external
force $F{\left(t\right)}$ which satisfies $F{\left(t\right)}=F{\left(t+\frac{2\pi}{\omega}\right)}$
and $F{\left(t\right)}=-F{\left(t+\frac{\pi}{\omega}\right)}$. Additionally,
the damping is introduced through the non-unit restitution coefficient
$0<e<1$. Similarly to the Hamiltonian case, we look for the periodic
solution, thus the impacts can be taken into account in the same form as above.
The solution should obey the following set of equations:

\begin{eqnarray}
\begin{array}{c}
\ddot{v}_{0}+\gamma_{1}v_{0}+\gamma_{2}\left(2v_{0}-v_{1}-v_{N}\right)=F{\left(t\right)}+\\
2p_{0}\delta_{0k}\underset{j=-\infty}{\overset{\infty}{\sum}}{\left(\begin{array}{c}
\delta{\left(t-\phi_{0}-\frac{\pi\left(2j+1\right)}{\omega}\right)}-\\
-\delta{\left(t-\phi_{0}-\frac{2\pi j}{\omega}\right)}
\end{array}\right)}
\end{array}\label{eq:25}\\
\begin{array}{c}
\ddot{v}_{n}+\gamma_{1}v_{n}+\gamma_{2}\left(2v_{k}-v_{k+1}-v_{k-1}\right)=F{\left(t\right)}+\\
2p_{k}\delta_{nk}\underset{j=-\infty}{\overset{\infty}{\sum}}{\left(\begin{array}{c}
\delta{\left(t-\phi_{k}-\frac{\pi\left(2j+1\right)}{\omega}\right)}-\\
-\delta{\left(t-\phi_{k}-\frac{2\pi j}{\omega}\right)}
\end{array}\right)}
\end{array}\\
\ddot{v}_{N}+\gamma_{1}v_{N}+\gamma_{2}\left(2v_{N}-v_{0}-v_{N-1}\right)=F{\left(t\right)}
\end{eqnarray}
where $\phi_{k}$ is the phase of the $k$-th particle with
respect to the external force $F{\left(t\right)}$.

The external force $F{\left(t\right)}$ can be removed from the equations
with the help of a simple transformation. Let $v_{n}{\left(t\right)}=u_{n}{\left(t\right)}+G{\left(t\right)}$
where $\ddot{G}{\left(t\right)}+\gamma_{1}G{\left(t\right)}=F{\left(t\right)}$.
Substitution into the above equations yields:
\begin{eqnarray}
\begin{array}{c}
\ddot{u}_{0}+\gamma_{1}u_{0}+\gamma_{2}\left(2u_{0}-u_{1}-u_{N}\right)=\\
=2p_{0}\delta_{0k}\underset{j=-\infty}{\overset{\infty}{\sum}}{\left(\begin{array}{c}
\delta{\left(t-\phi_{0}-\frac{\pi\left(2j+1\right)}{\omega}\right)}-\\
-\delta{\left(t-\phi_{0}-\frac{2\pi j}{\omega}\right)}
\end{array}\right)}
\end{array}\label{eq:28}\\
\begin{array}{c}
\ddot{u}_{n}+\gamma_{1}u_{n}+\gamma_{2}\left(2u_{n}-u_{n+1}-u_{n-1}\right)=\\
=2p_{k}\delta_{nk}\underset{j=-\infty}{\overset{\infty}{\sum}}{\left(\begin{array}{c}
\delta{\left(t-\phi_{k}-\frac{\pi\left(2j+1\right)}{\omega}\right)}-\\
-\delta{\left(t-\phi_{k}-\frac{2\pi j}{\omega}\right)}
\end{array}\right)}
\end{array}\label{eq:29}\\
\ddot{u}_{N}+\gamma_{1}u_{N}+\gamma_{2}\left(2u_{N}-u_{0}-u_{N-1}\right)=0\label{eq:30}
\end{eqnarray}

One can observe that the above equations are identical to those of the Hamiltonian
model, and therefore can be solved in a similar manner. Replacing
the impact terms with appropriate generalized Fourier series yields:
\begin{eqnarray}
\begin{array}{c}
\ddot{u}_{0}+\gamma_{1}u_{0}+\gamma_{2}\left(2u_{0}-u_{1}-u_{N}\right)=\\
=-\frac{4\omega p_{0}\delta_{0k}}{\pi}\underset{j=0}{\overset{\infty}{\sum}}{\cos{\left(\left(2j+1\right)\omega\left(t-\phi_{0}\right)\right)}}
\end{array}\label{eq:31}\\
\begin{array}{c}
\ddot{u}_{n}+\gamma_{1}u_{n}+\gamma_{2}\left(2u_{n}-u_{n+1}-u_{n-1}\right)=\\
=-\frac{4\omega p_{k}\delta_{nk}}{\pi}\underset{j=0}{\overset{\infty}{\sum}}{\cos{\left(\left(2j+1\right)\omega\left(t-\phi_{k}\right)\right)}}
\end{array}\label{eq:32}\\
\ddot{u}_{N}+\gamma_{1}u_{N}+\gamma_{2}\left(2u_{N}-u_{0}-u_{N-1}\right)=0\label{eq:33}
\end{eqnarray}

The ansatz has to be somewhat modified due to the phase differences:
\begin{equation}
u_{n}=\sum_{k}{\sum_{j=0}^{\infty}{u_{n,j,k}\cos{\left(\left(2j+1\right)\omega\left(t-\phi_{k}\right)\right)}}}\label{eq:34}
\end{equation}
where $u_{n,j,k}$ is derived in a way similar to the Hamiltonian
case:
\begin{equation}
\begin{array}{c}
u_{n,j,k}=\cfrac{4\omega p_{k}}{\pi\gamma_{2}\left(f_{j}-f_{j}^{-1}\right)\left(f_{j}^{-N-1}-1\right)}\times\\
\,\,\,\,\,\,\,\,\,\,\times\left(f_{j}^{\left|n-k\right|-N-1}+f_{j}^{-\left|n-k\right|}\right)
\end{array}
\end{equation}

Unlike the Hamiltonial model, we have here two sets of unknowns --
$p_{k}$ and $\phi_{k}$. Therefore, additional set of equations is
required. The first set is derived from the location of the barriers, as in the Hamiltonian case.
The mass $k_{s}\in k$ hits the right barrier at $t=\phi_{k_{s}}$:
\begin{equation}
\begin{array}{c}
v_{k_{s}}{\left(\phi_{k_{s}}\right)}=\sum_{k}{\frac{4\omega p_{k}}{\pi\gamma_{2}}\sum_{j=0}^{\infty}{\cfrac{\left(f_{j}^{\left|k_{s}-k\right|-N-1}+f_{j}^{-\left|k_{s}-k\right|}\right)}{\left(f_{j}-f_{j}^{-1}\right)\left(f_{j}^{-N-1}-1\right)}\times}}\\
\times\cos{\left(\left(2j+1\right)\omega\left(\phi_{k_{s}}-\phi_{k}\right)\right)}+G{\left(\phi_{k_{s}}\right)}=1
\end{array}
\end{equation}

In terms of the transformed variables, i.e. in terms of $u_{n}$,
the impacts should be symmetrical; therefore, one obtains $\dot{u}{\left(\phi_{k}^{-}\right)}=-\dot{u}{\left(\phi_{k}^{+}\right)}=p_{k}$.
However, in terms of the initial variables, the impact conditions \eqref{eq:7}
with the non-unit restitution coefficient should be satisfied. Thus,
one obtains the second set of equations:
\begin{equation}
\begin{array}{c}
\dot{v}{\left(\phi_{k}^{+}\right)}=\dot{u}{\left(\phi_{k}^{+}\right)}+\dot{G}{\left(\phi_{k}\right)}=\\
=-p_{k}+\dot{G}{\left(\phi_{k}\right)}=-e\left(p_{k}+\dot{G}{\left(\phi_{k}\right)}\right)=\\
=-e\left(\dot{u}{\left(\phi_{k}^{-}\right)}+\dot{G}{\left(\phi_{k}\right)}\right)=-e\dot{v}{\left(\phi_{k}^{-}\right)}
\end{array}
\end{equation}

\begin{equation}
\dot{G}{\left(\phi_{k_{s}}\right)}=qp_{k_{s}}
\end{equation}
where $q=\cfrac{1-e}{1+e}$. Final set of equations for the unknown
parameters of multi-breather solution is written as:
\begin{eqnarray}
\begin{array}{c}
G{\left(\phi_{k_{s}}\right)}=1-\,\,\,\,\,\,\,\,\,\,\,\,\,\,\,\,\,\,\,\,\,\,\,\,\,\,\,\,\,\,\\
-\sum_{k}{\frac{4\omega p_{k}}{\pi\gamma_{2}}\sum_{j=0}^{\infty}{\cfrac{\left(f_{j}^{\left|k_{s}-k\right|-N-1}+f_{j}^{-\left|k_{s}-k\right|}\right)}{\left(f_{j}-f_{j}^{-1}\right)\left(f_{j}^{-N-1}-1\right)}\times}}\\
\,\,\,\,\,\,\,\,\,\,\,\,\,\,\,\,\,\,\,\,\,\,\,\,\,\times\cos{\left(\left(2j+1\right)\omega\left(\phi_{k_{s}}-\phi_{k}\right)\right)}
\end{array}\label{eq:39}\\
\dot{G}{\left(\phi_{k_{s}}\right)}=qp_{k_{S}}\,\,\,\,\,\,\,\,\,\,\,\,\,\,\,\,\,\,\,\,\,\,\,\,\,\,\,\,\,\,\,\,\,\,\,\,\,\,\,\,\,\label{eq:40}
\end{eqnarray}

Note that $\phi_{k}$ appears in the equations in a way that does
not allow exact solution. So, additional simplifications or numeric
approaches are required after this point.

\subsubsection{In-Phase DBs}

Simplification of equation \eqref{eq:39} is possible, if one considers
the multi-breather with all particles having the same phase with respect
to the external forcing. This is a special case $\phi_{k}=\phi$.
With this assumption, equations \eqref{eq:39} become much simpler,
since $\phi$ vanishes from the summations and equations \eqref{eq:39}-\eqref{eq:40}
are reduced to the following form:
\begin{eqnarray}
G{\left(\phi\right)} & = & 1-\sum_{k}{\frac{4\omega p_{k}}{\pi\gamma_{2}}\chi_{k_{s},k}}\label{eq:41}\\
\dot{G}{\left(\phi\right)} & = & qp_{k_{s}}\label{eq:42}
\end{eqnarray}
The second equation demands that $p_{k}=p$. Furthermore, the first
equation then gives $\sum_{k}{\chi_{k_{s},k}}\equiv\sigma$ regardless
of $k_{s}$ which means that the two sets of equations are independent
of $k_{s}$. This leads to two conclusions. The first is that we only
have to solve a set of two equations. The second one is that the forced
in-phase DBs are only possible when certain symmetries are satisfied
-- in order for $\sum_{k}{\chi_{k_{s},k}}$ to be the same for any
$k_{s}$ , the proximity of each localization site to all other sites
should be the same for all sites. For 2-site DB this is always true;
however, for 3 and more sites this is only true if they are equally
spaced, with account of the periodic boundary conditions.

Simple choice of harmonic forcing $F=A\cos{\left(\omega t\right)}$
yields:
\begin{equation}
G{\left(t\right)}=\tilde{A}\cos{\left(\omega t\right)}
\end{equation}

where $\tilde{A}=-\frac{A}{\omega^{2}-\gamma_{1}}$.

Substitution of \eqref{eq:41} and \eqref{eq:42} then leads to the
following system of equations:

\begin{eqnarray}
\tilde{A}\cos{\left(\omega\phi\right)} & = & 1-\frac{4\omega p}{\pi\gamma_{2}}\sigma\label{eq:44}\\
-\tilde{A}\omega\sin{\left(\omega\phi\right)} & = & qp\label{eq:45}
\end{eqnarray}

Possible values of $p$ are easily obtained:
\begin{equation}
p=\cfrac{4\pi\gamma_{2}\omega^{3}\sigma\pm\pi\gamma_{2}\omega\sqrt{\left(4\omega^{2}\sigma\right)^{2}\tilde{A}^{2}+\left(\pi\gamma_{2}q\right)^{2}\left(1-\tilde{A}^{2}\right)}}{\left(\left(4\omega^{2}\sigma\right)^{2}+\left(q\pi\gamma_{2}\right)^{2}\right)}
\end{equation}

This solution can be plugged back into equations \eqref{eq:44}-\eqref{eq:45}
to determine which of them is physically meaningful, and to obtain the value of $\text{\ensuremath{\phi}}$.

\section{\label{sec: IV}Stability}

The stability of the periodic multi-breather solutions will be investigated
with the help of Floquet theory\citep{floquet}. The Floquet multipliers
are often evaluated numerically, but as mentioned in the introduction,
the explored model allows explicit construction of the monodromy matrix.
Then, it is easy to find its eigenvalues for every set of parameters;
thus, broad regions of the parameter space can be explored for various
structures of the breathers, and with limited numeric efforts. Moreover,
the eigenvectors corresponding to the unstable Floquet multipliers
can be easily computed and examined to give a qualitative insight
into physical mechanisms of the loss of stability.

The governing equations of motion for Hamiltonian model can be re-written
in the following equivalent form:
\begin{equation}
\dot{\vec{u}}=\mbox{A}\vec{u}
\end{equation}

where $\vec{u}=\left[\begin{array}{cccccc}
u_{0} & \cdots & u_{N} & \dot{u}_{0} & \cdots & \dot{u}_{N}\end{array}\right]^{T}$ and:
\begin{equation}
\mbox{A}=\left[\begin{array}{cc}
\mbox{0}_{\left(N+1\right)\times\left(N+1\right)} & \mbox{I}_{\left(N+1\right)\times\left(N+1\right)}\\
\tilde{\mbox{A}}_{\left(N+1\right)\times\left(N+1\right)} & \mbox{0}_{\left(N+1\right)\times\left(N+1\right)}
\end{array}\right]
\end{equation}
\begin{equation}
\begin{array}{c}
\mbox{\ensuremath{\tilde{A}}}=\,\,\,\,\,\,\,\,\,\,\,\,\,\,\,\,\,\,\,\,\,\,\,\,\,\,\,\,\,\,\,\,\,\,\,\,\,\,\,\,\,\,\,\,\,\,\,\,\,\,\,\,\,\,\,\,\,\,\,\,\,\,\,\,\,\,\,\,\,\,\,\,\,\,\,\,\\
\left[\begin{array}{cccccc}
\gamma_{1}+2\gamma_{2} & -\gamma_{2} & 0 & \cdots & 0 & -\gamma_{2}\\
-\gamma_{2} & \gamma_{1}+2\gamma_{2} & -\gamma_{2} & 0 & \cdots & 0\\
0 & -\gamma_{2} & \ddots & \ddots & \ddots & \vdots\\
\vdots & \ddots & \ddots & \gamma_{1}+2\gamma_{2} & -\gamma_{2} & 0\\
0 & \cdots & 0 & -\gamma_{2} & \gamma_{1}+2\gamma_{2} & -\gamma_{2}\\
-\gamma_{2} & 0 & \cdots & 0 & -\gamma_{2} & \gamma_{1}+2\gamma_{2}
\end{array}\right]
\end{array}
\end{equation}

Here $\tilde{A}$ is Laplace adjacency matrix of the system. For the forced-damped model, minor modification is required:
\begin{equation}
\dot{\vec{v}}=\mbox{A}\vec{v}+\vec{F}
\end{equation}
where $\vec{F}=F{\left(t\right)}\left[\begin{array}{ccc}
0 & \cdots & 0\end{array}\begin{array}{ccc}
1 & \cdots & 1\end{array}\right]^{T}$.

All considered solutions, both for Hamiltonian and forced damped system,
are symmetric in a sense that the successive impacts for each particle
are divided by half-period intervals, and absolute amounts of momentum transferred to given particle in the course of given impact is the same.
From the above equation, it
is easy to derive the matrix, that describes the evolution of perturbed
phase trajectory between two successive impacts:
\begin{equation}
\mbox{L}=\exp\left(\cfrac{\pi}{\omega}A\right)
\end{equation}

To describe the evolution of the perturbed phase trajectory in the
course of impacts, we apply a formalism of saltation matrix \citep{saltation}.
Since the impacts are instantaneous independent events, they can be
treated separately and then combined to result in following saltation
matrix:
\begin{equation}
\mbox{S}=\left[\begin{array}{cc}
\mbox{\ensuremath{\tilde{S}}}_{\left(N+1\right)\times\left(N+1\right)} & \mbox{0}_{\left(N+1\right)\times\left(N+1\right)}\\
\hat{\mbox{S}}_{\left(N+1\right)\times\left(N+1\right)} & \mbox{\ensuremath{\tilde{S}}}_{\left(N+1\right)\times\left(N+1\right)}
\end{array}\right]
\end{equation}
where
\begin{widetext}
{\small{}
\begin{equation}
\tilde{\mbox{S}}=\left[\begin{array}{ccccc}
1-\left(1+e\right)\sum_{k}{\delta_{1k}} & 0 & \cdots & \cdots & 0\\
0 & 1-\left(1+e\right)\sum_{k}{\delta_{2k}} & 0 &  & \vdots\\
\vdots & 0 & \ddots & \ddots & \vdots\\
\vdots & \mbox{} & \ddots & 1-\left(1+e\right)\sum_{k}{\delta_{\left(N-1\right)k}} & 0\\
0 & \cdots & \cdots & 0 & 1
\end{array}\right]\label{eq:53}
\end{equation}
\begin{equation}
\hat{\mbox{S}}=\left[\begin{array}{ccccc}
\cfrac{\left(1+e\right)\sum_{k}{\delta_{1k}\psi_{k}}}{\Gamma_{1}} & 0 & \cdots & \cdots & 0\\
0 & \cfrac{\left(1+e\right)\sum_{k}{\delta_{2k}\psi_{k}}}{\Gamma_{2}} & 0 & \mbox{} & \vdots\\
\vdots & 0 & \ddots & \ddots & \vdots\\
\vdots & \mbox{} & \ddots & \cfrac{\left(1+e\right)\sum_{k}{\delta_{\left(N-1\right)k}\psi_{k}}}{\Gamma_{N-1}} & 0\\
0 & \cdots & \cdots & 0 & 1
\end{array}\right]\label{eq:54}
\end{equation}
}{\small \par}

\noindent with $\psi_{k}=\ddot{u}_{k}{\left(\phi-\right)}$ and $\Gamma_{k}=p_{k}$
for the hamiltonian model, and $\psi_{k}=\ddot{v}_{k}{\left(\phi-\right)}$
and $\Gamma_{k}=\Gamma=p+\dot{G}\left(\phi\right)$ for the forced-damped
model. Note that for the Hamiltonian model the coefficient of restitution
$e$ is set to unity.
\end{widetext}

Due to the symmetry of even functions composing the Fourier series,
the monodromy matrix can be written compactly as follows:
\begin{equation}
\mbox{M}=\left(\mbox{L}\mbox{S}\right)^{2}
\end{equation}

As it was mentioned above, the eigenvalues of this monodromy matrix are computed numerically for given parameter values. The resulting stability pattern in the space of parameters are exemplified in the next section.

\section{Numeric Validation and Stability Patterns \label{sec:Numeric-Validation}}

\subsection{Hamiltonian Model}

\begin{figure}[tb]
\begin{centering}
\includegraphics[angle=0, width=0.75\linewidth]{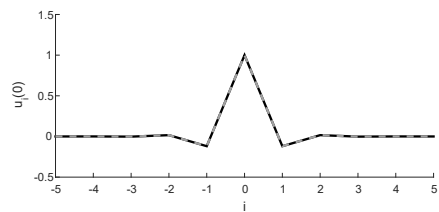}
\par\end{centering}

\protect\caption{\label{fig:1}Comparison of the displacement amplitude of the particles
between $N=10$ (black) and $N=300$ (dashed gray) with a single site
DB. Negative indices denotes the $i$-th mass in the chain
where $i=N+1+\left(\mbox{negative index}\right)$ to better represent
the periodic boundary condition.}
\end{figure}

In order to assess the properties of the analytic solution, and to
verify the accuracy of numeric algorithms, we compare the results
of the analysis to numeric simulations. The simulations were performed
in MatLab; the vibro-impact was modeled according to the impact law
using built-in event-driven algorithms with Runge-Kutta (RK) solver.
Simulations show that the analytic solutions perfectly coincide with
the numeric results. Fig. \ref{fig:1} demonstrates that, as one would
expect, if the DB is exponentially localized, it looks very similar
both for very short and very long chains.

\begin{figure}[tb]
\begin{centering}
\includegraphics[angle=0, width=0.75\linewidth]{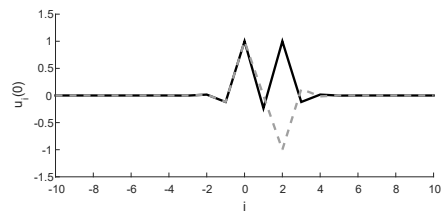}
\par\end{centering}

\protect\caption{\label{fig:2}Comparison of the displacement amplitude of the particles
between in-phase (black) and anti-phase (dashed gray) 2-site MB. Negative
indices denotes the $i$-th mass in the chain where $i=N+1+\left(\mbox{negative index}\right)$
to better represent the periodic boundary condition.}
\end{figure}

\begin{figure}[tb]
\begin{centering}
\includegraphics[angle=0, width=0.75\linewidth]{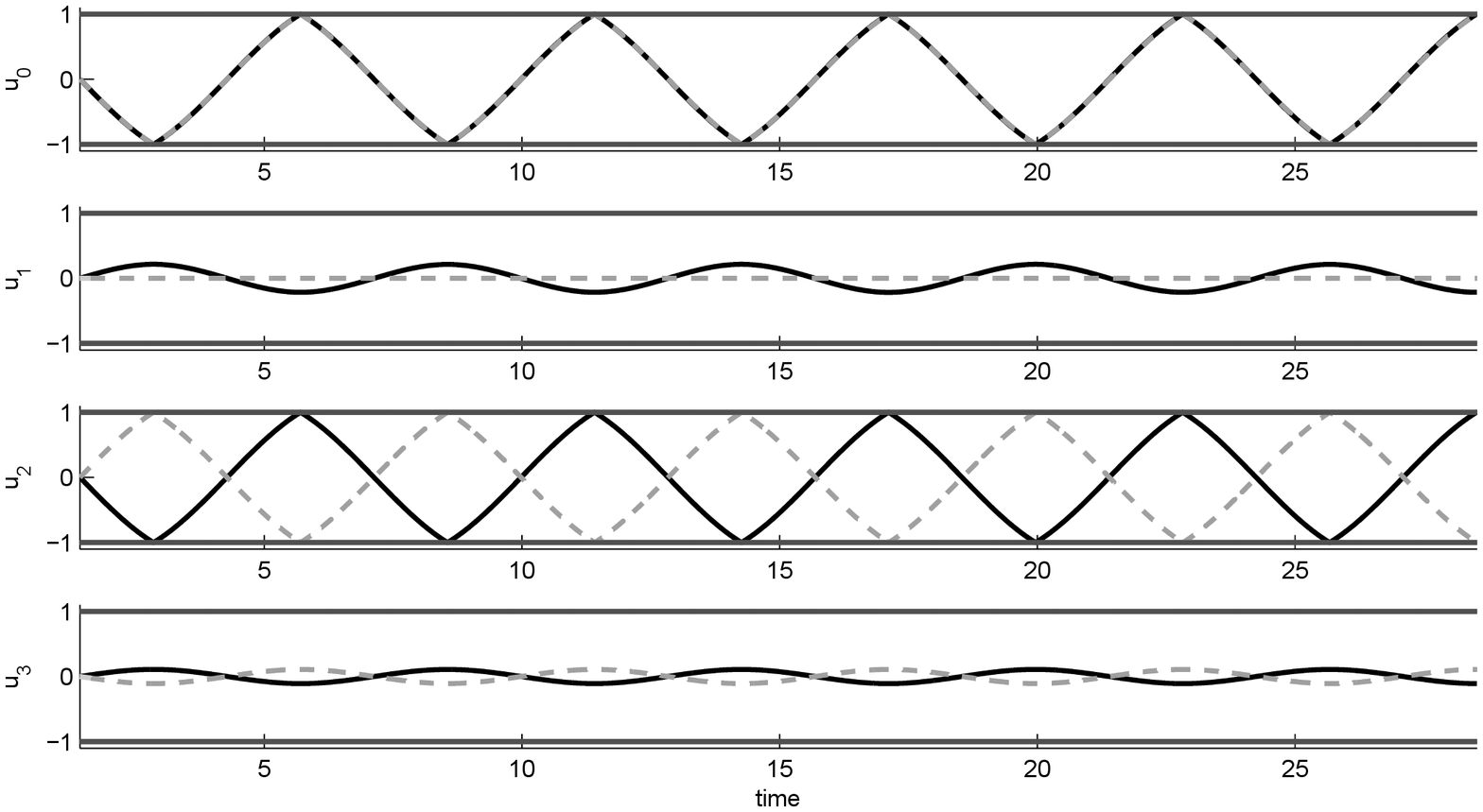}
\par\end{centering}

\protect\caption{\label{fig:3}Comparison of the displacements of the first 4 particles
between in-phase (black) and anti-phase (dashed gray) 2-site MB for
$N=20$.}
\end{figure}

When considering the MBs, there are two possible states
for each site -- in-phase or out-of-phase with respect to the $0$-th
mass. Both options in the case of 2-site DB are presented in Fig.
\ref{fig:2}. Due to the symmetry, when the two sites are in anti-phase,
the sum of the two forces applied to the mass between them is zero,
i.e the mass is in complete halt. Besides this phenomenon, the localization
appears to be similar; however, the other masses are in
opposite phases as seen in Fig. \ref{fig:3}.

\begin{figure}[tb]
\begin{centering}
\includegraphics[angle=0, width=0.75\linewidth]{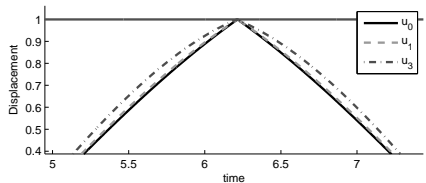}
\par\end{centering}

\protect\caption{\label{fig:4}Displacements of impacting particles for the 3-site MB with excited
sites at $0,1$ and $3$ for $N=20$.}
\end{figure}

While the MB has only two sites, the formation is necessarily
symmetric and therefore the displacement of the impacting masses remains
identical (or inverse if in anti-phase). When there are more than
two sites, this symmetry can be broken. The oscillations are still
in-phase (or anti-phase), but the impacts are not equivalent, i.e.
different particles exchange different amounts of momentum with the
constraints in the course of impacts, as demonstrated in Fig. \ref{fig:4}.

\subsection{Forced-Damped Model}

\begin{figure}[tb]
\begin{centering}
\includegraphics[angle=0, width=0.75\linewidth]{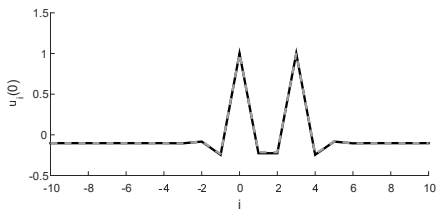}
\par\end{centering}

\protect\caption{\label{fig:5}Comparison of the displacement amplitude of the particles
between $N=20$ (dashed gray) and $N=300$ (black) with a 2-site MB.
Negative indices denote the $i$-th mass in the chain where
$i=N+1+\left(\mbox{negative index}\right)$ to better represent the
periodic boundary condition.}
\end{figure}

In the case of the forced-damped model, the numeric simulations
are in accordance with the analytic solution. It is also clear
that, unlike the Hamiltonian model where the amplitude rapidly converges
to zero as the particle is further away from the localization site,
in the forced-damped model it converges as rapidly, but to $G{\left(t\right)}$
instead. Fig. \ref{fig:5} illustrates the difference between the MBs
in similar chains with relatively large and small number of particles.

\begin{figure}[tb]
\begin{centering}
\includegraphics[angle=0, width=0.75\linewidth]{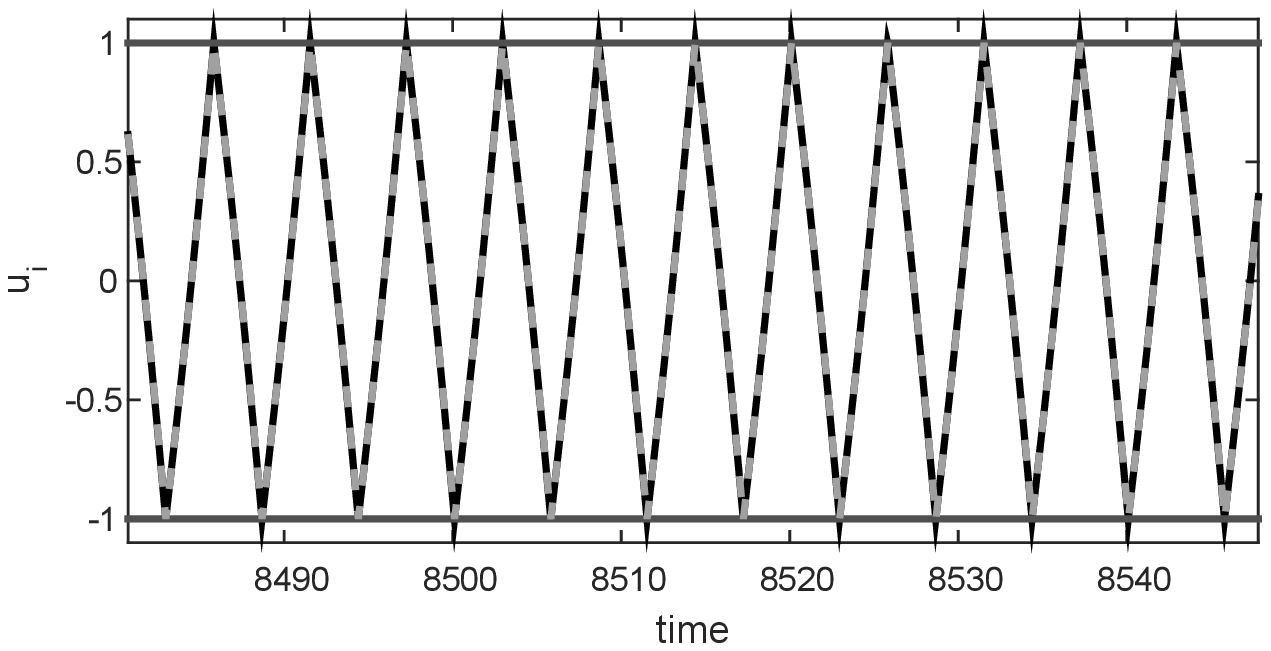}
\par\end{centering}

\begin{centering}
\includegraphics[angle=0, width=0.75\linewidth]{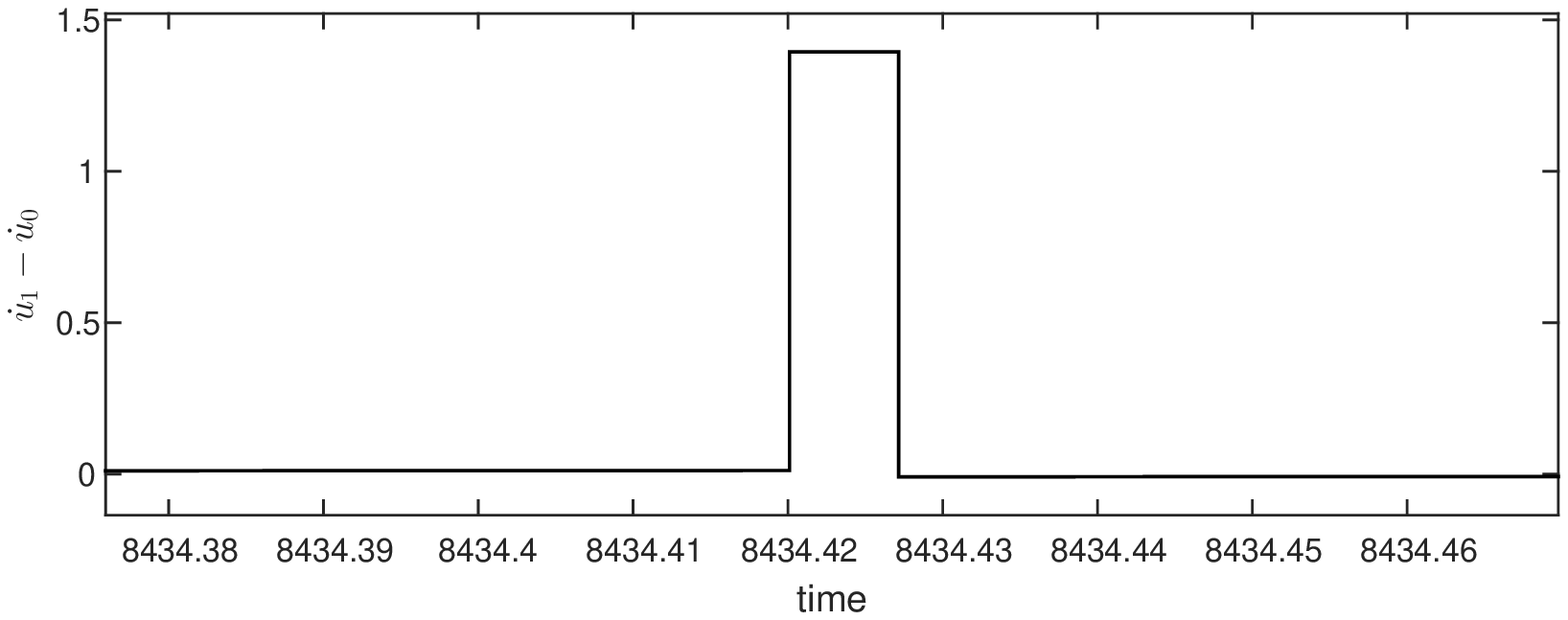}
\par\end{centering}

\protect\caption{\label{fig:6}Top: Displacements $u_{0}$ (black) and $u_{1}$ (dashed
gray) for a Forced 3-site MB with sites at $n=0,1,2$ with $N=20$.
Bottom: The difference in the velocities $\dot{u}_{1}-\dot{u}_{0}$.}
\end{figure}

It is difficult to obtain analytic results for the MBs with different phases at different sites, since these solutions are no more symmetric. Thus, they has been found numerically.
Fig. \ref{fig:6} shows a simple example of such solution. The difference in velocities clearly
indicates the phase differences between the localization sites.

\subsection{Stability}

\begin{figure}[tb]
\begin{centering}
\includegraphics[angle=0, width=0.75\linewidth]{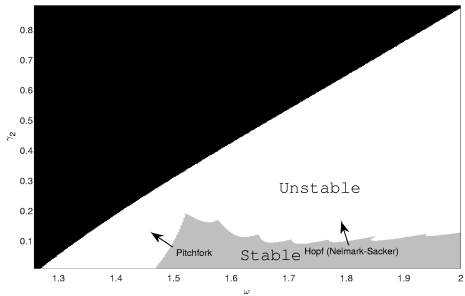}
\par\end{centering}

\protect\caption{\label{fig:7}Existence-stability map for a 2-site Forced MB with
sites at $0$ and $1$ for $N=20$, $\gamma_{1}=0.1$, $e=0.9$ and
$A=1.5$. In the black zone the DB solution with the considered structure does not exist.}
\end{figure}

\begin{figure}[tb]
\begin{centering}
\includegraphics[angle=0, width=0.75\linewidth]{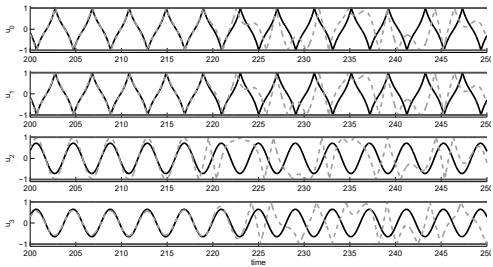}
\par\end{centering}

\protect\caption{\label{fig:8}Analytic prediction (black) and numerical approximation
(dashed gray) of the displacements of the first 4 masses for the unstable
forced DB.}
\end{figure}

The stability analysis can be more easily verified and illustrated
in the forced-damped case, since the solutions are hyperbolic dynamical attractors.
The analysis reveals two mechanisms of the loss of stability
-- via pitchfork bifurcation (corresponding to Floquet multipliers
leaving the unit circle through positive side of the real axis)
and Hopf (Neimark-Sacker) bifurcation which corresponds to a conjugate pair
of complex Floquet multipliers leaving the unit circle. Fig. \ref{fig:7}
shows the existence-stability map for the two-site multi-breather
in the plane of $\omega-\gamma_{2}$ where both possible bifurcation
scenarios are present. The MB solution ceases to exist if the frequency crosses the boundary of the propagation zone, or if some particle achieves the grazing limit. The latter restriction means that either the displacement of one of the "non-impacting" particles approaches unity, or the amount of momentum transferred to the "impacting" particle approaches zero. In both cases, the solution can remain localized, but should be re-derived due to modification of the subset of the "impactimg" particles. Fig. \ref{fig:8} presents the examples of the DBs for sets of parameters
in the stable and unstable zones.

\begin{figure}[tb]
\begin{centering}
\includegraphics[angle=0, width=0.75\linewidth]{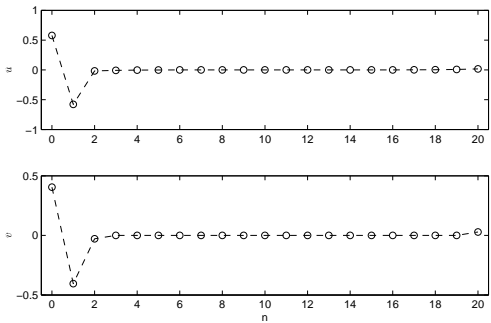}
\par\end{centering}

\protect\caption{\label{fig:9}Values of a typical eigenvector corresponding to the pitchfork
bifurcation for the 2-site forced MB with excitation sites at $0$ and $1$. $u$
and $v$ denote displacements and velocity components of the eigenvectors
respectively.}
\end{figure}

\begin{figure}[tb]
\begin{centering}
\includegraphics[angle=0, width=0.75\linewidth]{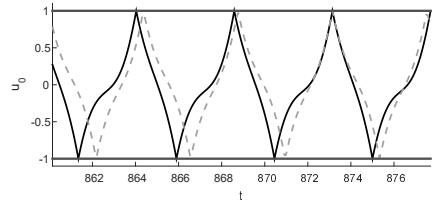}
\par\end{centering}

\protect\caption{\label{fig:10}Analytic prediction for the symmetric solution (dashed
gray) and numerical simulation (black) of the displacement of the
impacting particle for $N=20$ and $\gamma_{2}=0.07$ with the 2-site forced
MB after breaking of the symmetry.}
\end{figure}

As it was mentioned before, there are two mechanisms for the loss of stability.
One can gain some insight into physical reasons for the loss of stability
by inspection of the corresponding eigenvectors of the
monodromy matrix. For the pitchfork bifurcation, Fig. \ref{fig:9}
shows strongly localized eigenvector. Furthermore, the localization
is at the MB sites and appears to be anti-symmetric. This means that
the bifurcation leads to breaking of the symmetry. By slow "sweeping"
the frequency from the stable regime to the unstable, the new branch
created by the pitchfork bifurcation can be traced, and the asymmetric
MB appears as presented in Fig. \ref{fig:10}

\begin{figure}[tb]
\begin{centering}
\includegraphics[angle=0, width=0.75\linewidth]{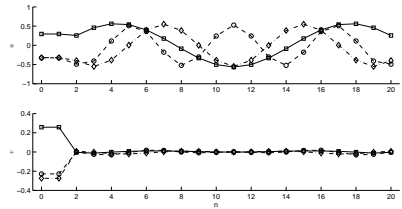}
\par\end{centering}

\protect\caption{\label{fig:11} Spatial profiles of several eigenvectors corresponding
to Neimark-Saker bifurcation for the 2-site forced MB with sites at $0$ and
$1$. $u$ and $v$ denote displacements and velocity components of
the eigenvectors respectively. Square marker, diamond marker and circle
marker correspond to the first, the second and the third ``well'' from the
left respectively.}
\end{figure}

\begin{figure}[tb]

\begin{center}
\includegraphics[angle=0, width=0.75\linewidth]{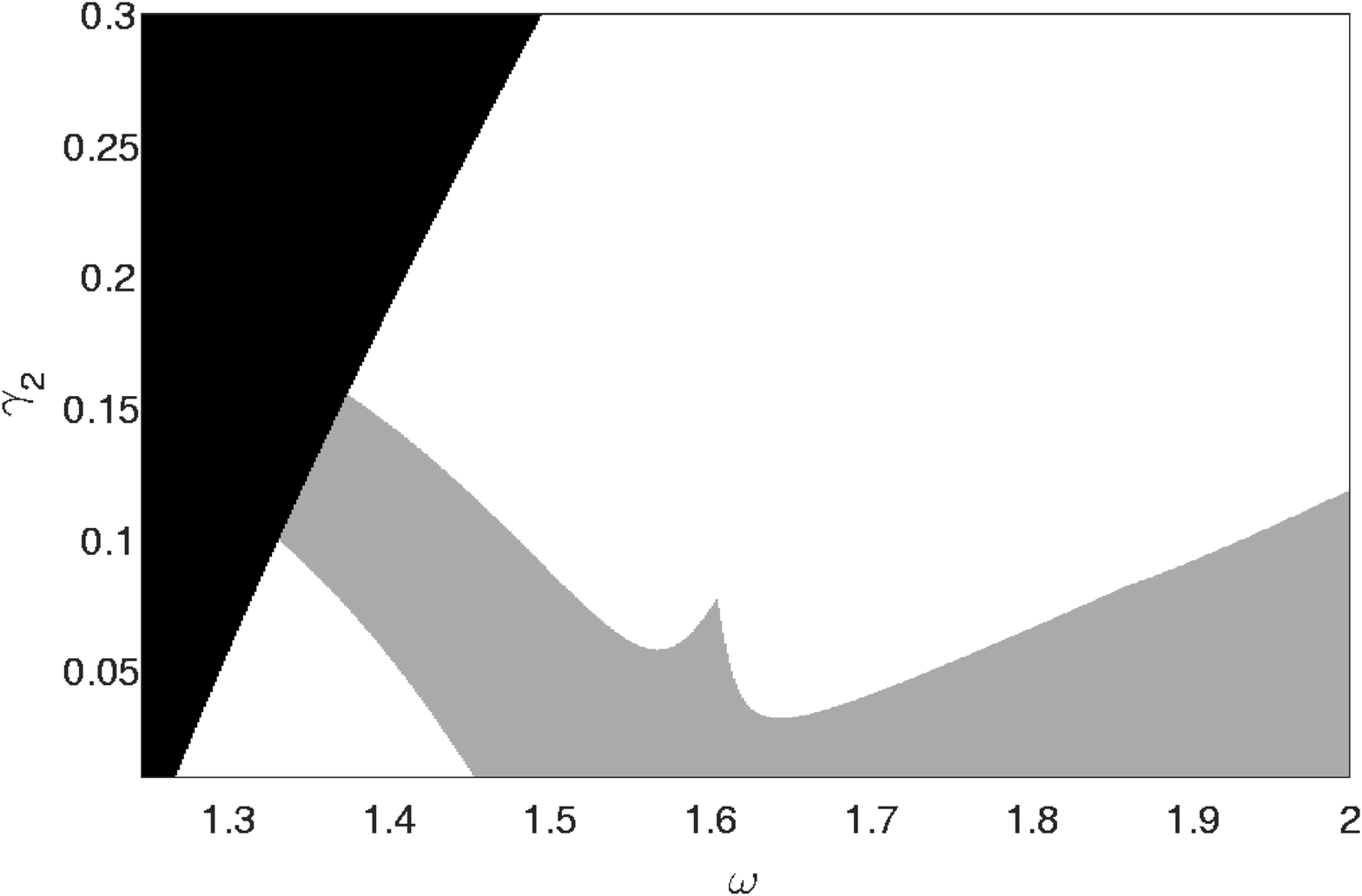}
\par\end{center}%

\begin{center}
\includegraphics[angle=0, width=0.75\linewidth]{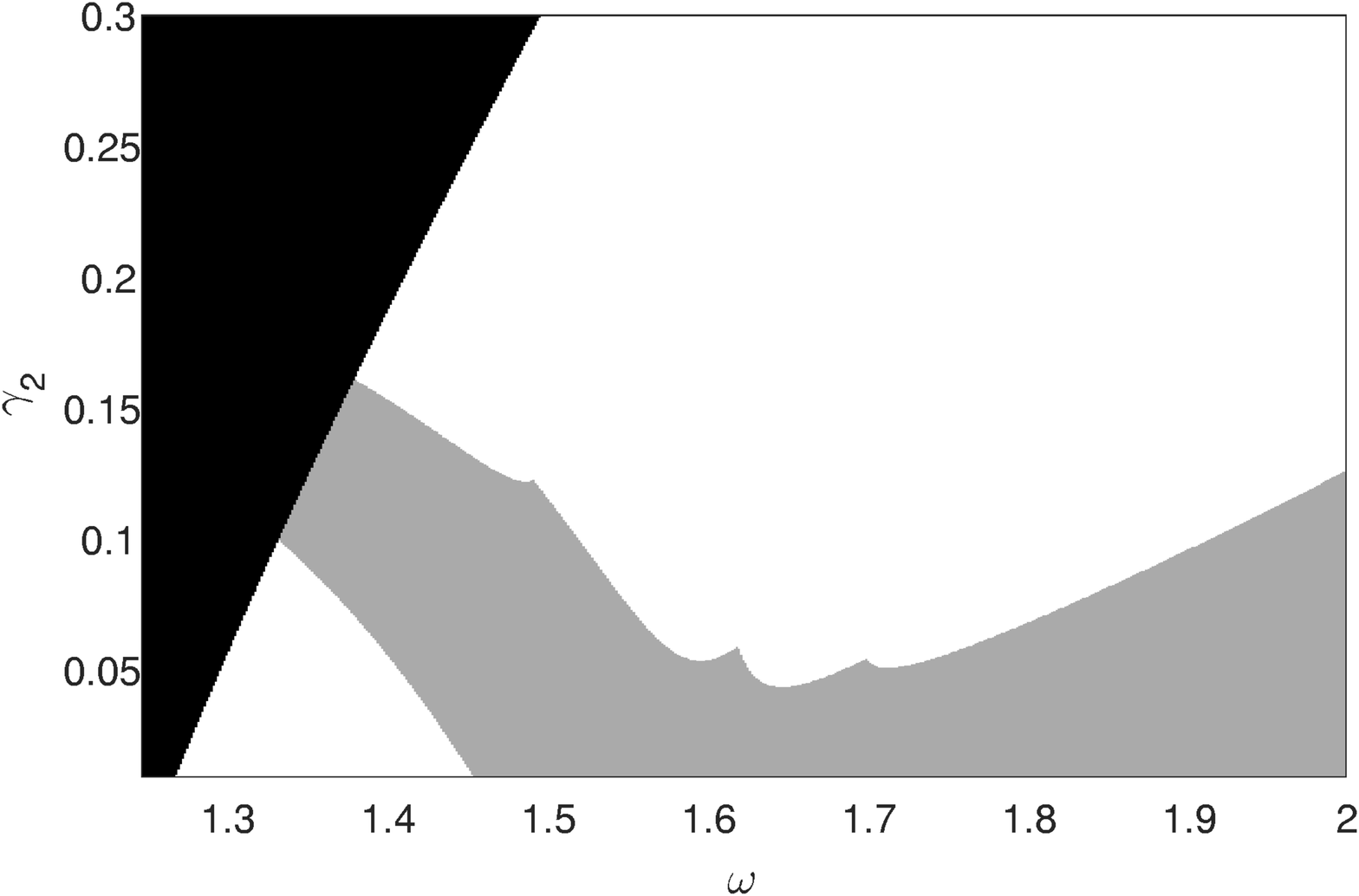}
\par\end{center}%

\begin{center}
\includegraphics[angle=0, width=0.75\linewidth]{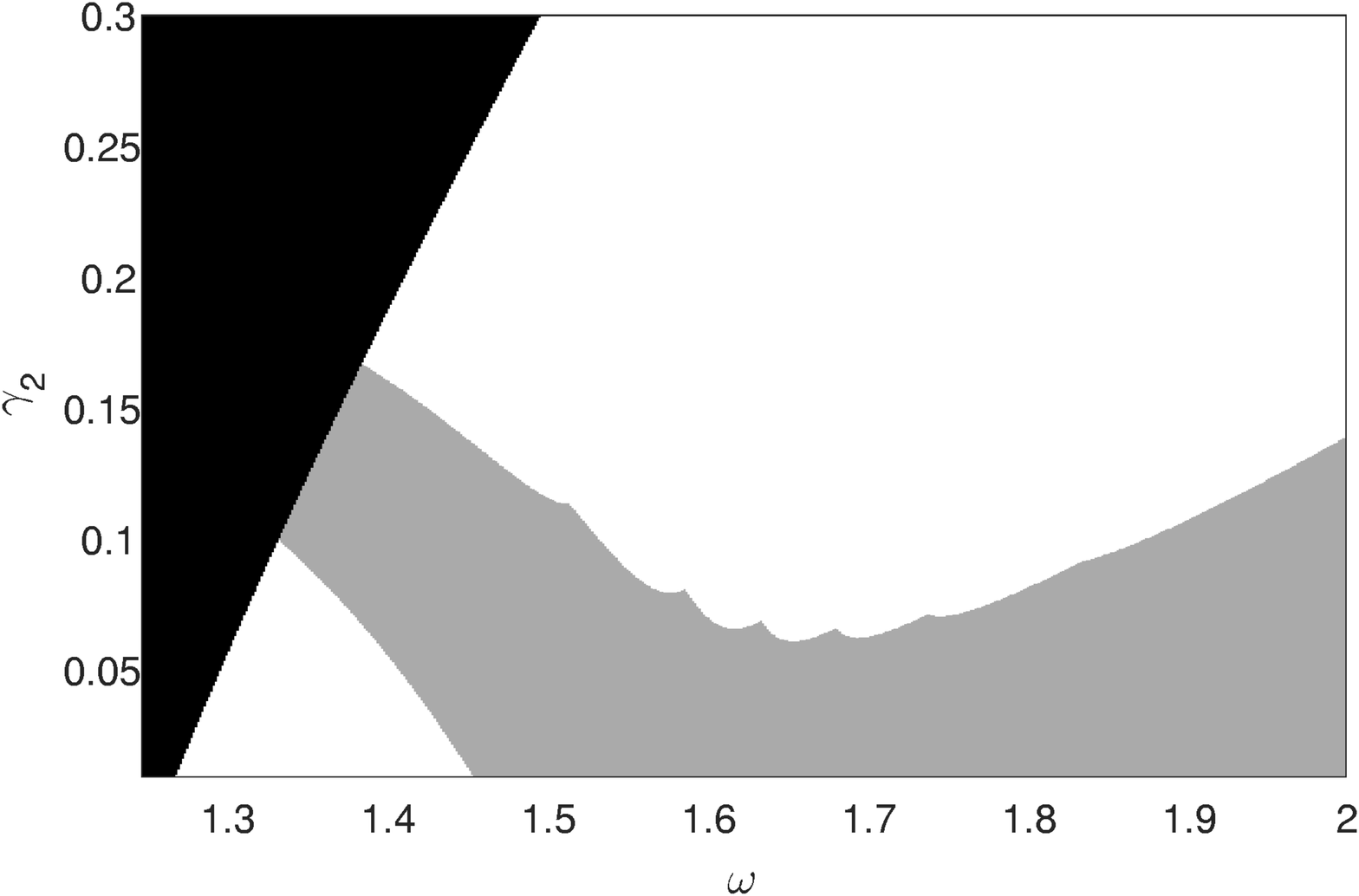}
\par\end{center}%

\begin{center}
\includegraphics[angle=0, width=0.75\linewidth]{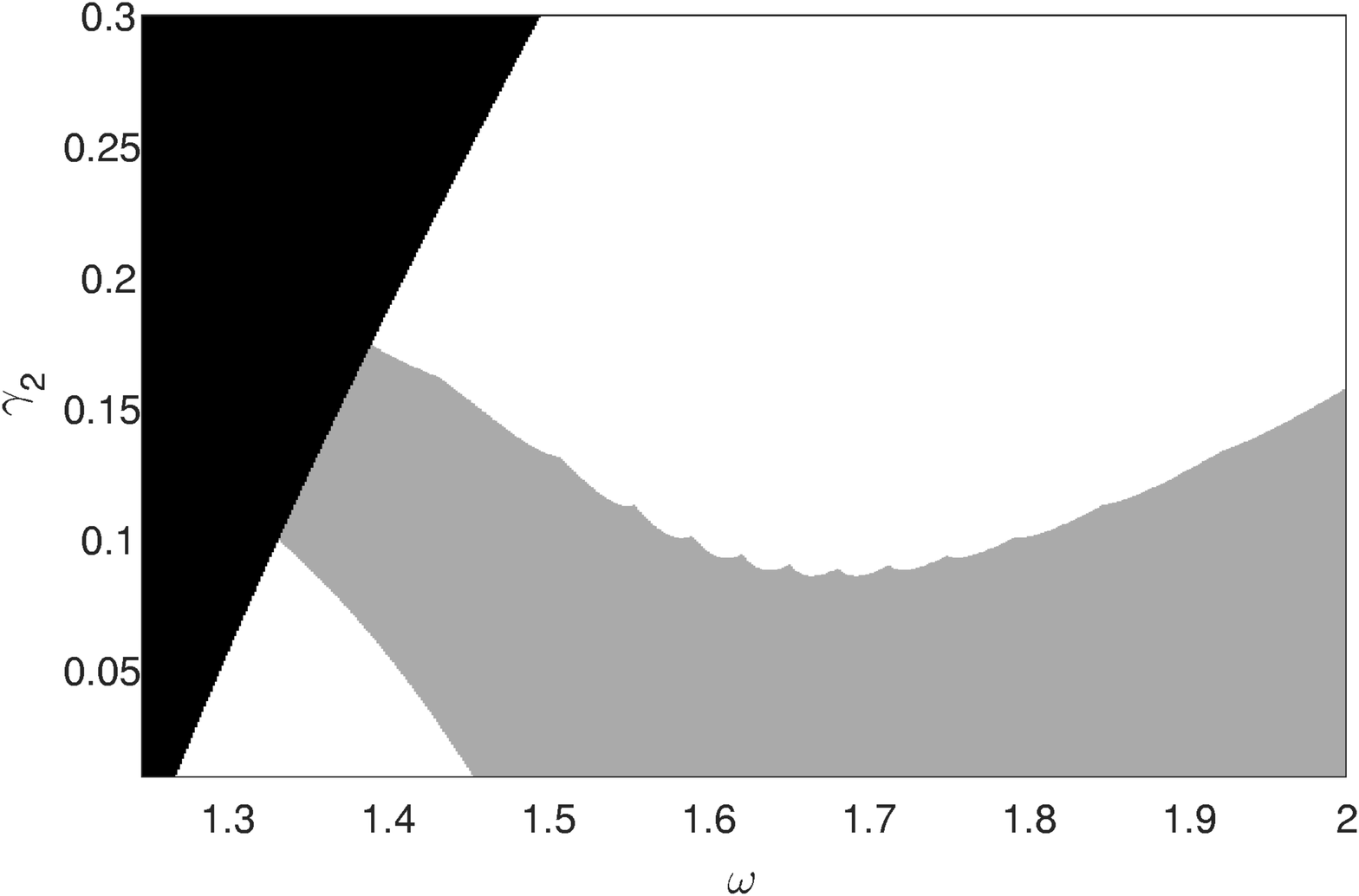}
\par\end{center}%

\protect\caption{\label{fig:12}Stability map for the single-site DB with $\gamma_{1}=0.1$,
$e=0.9$ and $A=1.5$ for (a) $N=5$ (b) $N=10$ (c) $N=20$ (d) $N=40$(top-down)}
\end{figure}

For the case of Neimark - Saker bifurcation, Fig. \ref{fig:7} demonstrates
peculiar structure for the stability boundary with multiple ``wells''.
Computation of the eigenvectors responsible for the loss of stability on the boundaries of these wells reveals that each well is related
to different spatial mode of the loss of stability. Examples of these modes are presented in Fig. \ref{fig:11}.
While the part corresponding to the velocity is mostly localized at the MB sites, the part corresponding to the displacements is not localized, and apparently depends on the number of particles in the chain fragment.

From Fig. \ref{fig:11} it is clear, that sharp differences between
the neighboring wells occur due to the fact that the chain is
relatively short, so there is big difference between lower eigenmodes.
So, one can conjecture profound "well structure" for the stability
boundary of the MBs in relatively short chains, and transition to
smooth boundaries for longer chains. Fig. \ref{fig:12} confirms this
conjecture and reveals clear correlation between the number of particles
and the number of the wells. It also comes to explain why this peculiar
structure was not observed in previous works. The wells become
smaller and more dense as the number of particles increases. It is
also interesting to note that the structure of the ``pitchfork''
fragment of the stability boundary does not seem to be significantly
affected by the number of particles. This is understandable, since
the corresponding eigenvector in all explored cases is strongly localized
at the DB site.

\begin{figure}[tb]
\begin{centering}
\includegraphics[angle=0, width=0.75\linewidth]{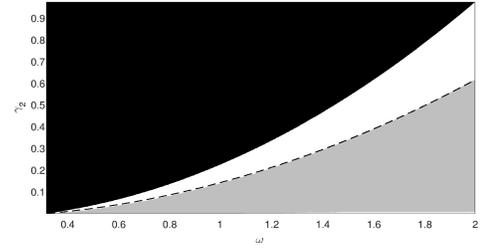}
\par\end{centering}

\protect\caption{\label{fig:13}Existence - stability map for the single-site conservative
DB for $N=20$ and $\gamma_{1}=0.1$. The dashed line corresponds
to $\omega^{2}=\gamma_{1}+\left(19/3\right)\gamma_{2}$}
\end{figure}

The stability of the Hamiltonian model is more difficult for numeric verification,
since the solutions are not attractors. However, interesting data
may be procured from it, especially with respect to appearance of
unstable Floquet multipliers with non-zero imaginary part. In Fig.
\ref{fig:13} the loss of stability occurs through the Neimark-Saker bifurcation is shown
together with the approximation of the border line. One can thus conjecture
that the Hopf bifurcation occurs due to interaction with the boundary of the propagation zone.

\begin{figure}[tb]
\begin{centering}
\includegraphics[angle=0, width=0.75\linewidth]{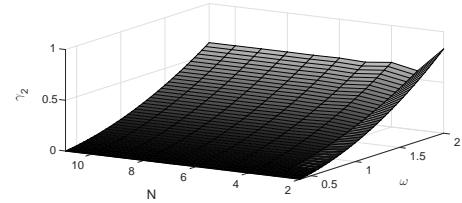}
\par\end{centering}

\protect\caption{\label{fig:14}Evolution of the stability boundary of the single-site
DB in the frequency-stiffness domain as a function of the system size.}
\end{figure}

\begin{figure}[tb]
\begin{centering}
\includegraphics[angle=0, width=0.75\linewidth]{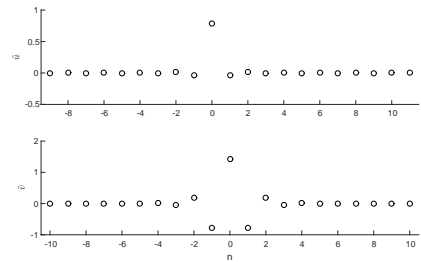}
\par\end{centering}

\protect\caption{\label{fig:15} Shape of eigenvector corresponding to the Neimark-Saker bifurcation
for the single-site conservative DB shown in Fig. \ref{fig:13}. $\tilde{u}$
and $\tilde{v}$ denote displacements and velocity components of the
eigenvector respectively. }
\end{figure}

The analytic result shows that the solution converges to that of the
infinite chain as the size of the system grows. The numerics shows
that the existence-stability map appears to be very similar regardless
 the system size. In Fig. \ref{fig:14} we demonstrate that any noticeable
modifications of the stability boundary occur only for extremely short
chains with $N=3-4$. This is very different from the stability patterns
observed for the forced-damped case. Corresponding eigenvector turns
out to be strongly localized, as presented in Fig. \ref{fig:15}.

\begin{figure}[tb]
\begin{centering}
\includegraphics[angle=0, width=0.75\linewidth]{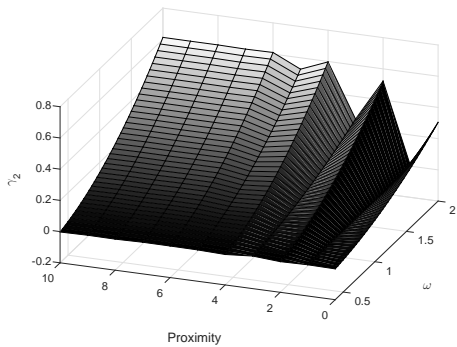}
\par\end{centering}

\protect\caption{\label{fig:16}Evolution of the loss of stability boundary of the 2-site
MB in the frequency-stiffness domain as a function of distance between the excited 
sites (value $0$ corresponds to two consecutive excited sites).}
\end{figure}

In addition to the system size, it is also interesting to examine the
effect of the proximity between the localization sites for the MB solution. One would expect
its influence to vanish very quickly as the distance between the sites increases,  due to
the strong localization; it is indeed the case as shown in Fig. \ref{fig:16}.
However, one should note that the proximity has a very strong influence
when the localization sites are close.

\begin{figure}[tb]
\begin{centering}
\includegraphics[angle=0, width=0.75\linewidth]{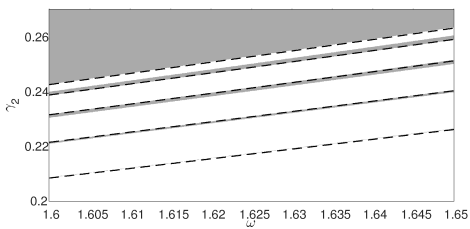}
\par\end{centering}

\protect\caption{\label{fig:17}Stability map for the 2-site MB with excitation sites at 0 and 2
in anti-phase for $N=20$ and $\gamma_{1}=0.1$.}
\end{figure}

\begin{figure*}[tb]
\begin{centering}
\includegraphics[width=0.4\linewidth]{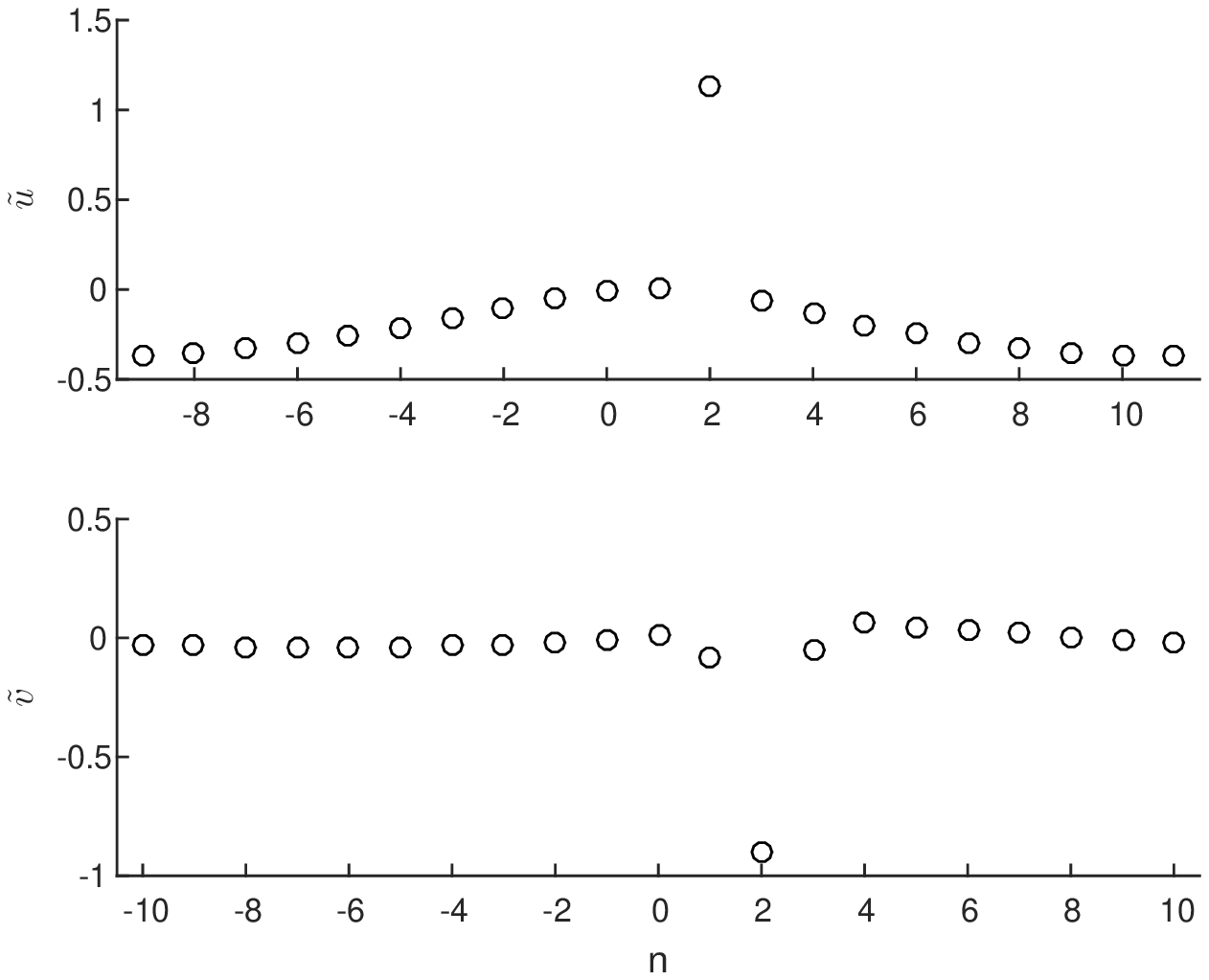}\includegraphics[width=0.4\linewidth]{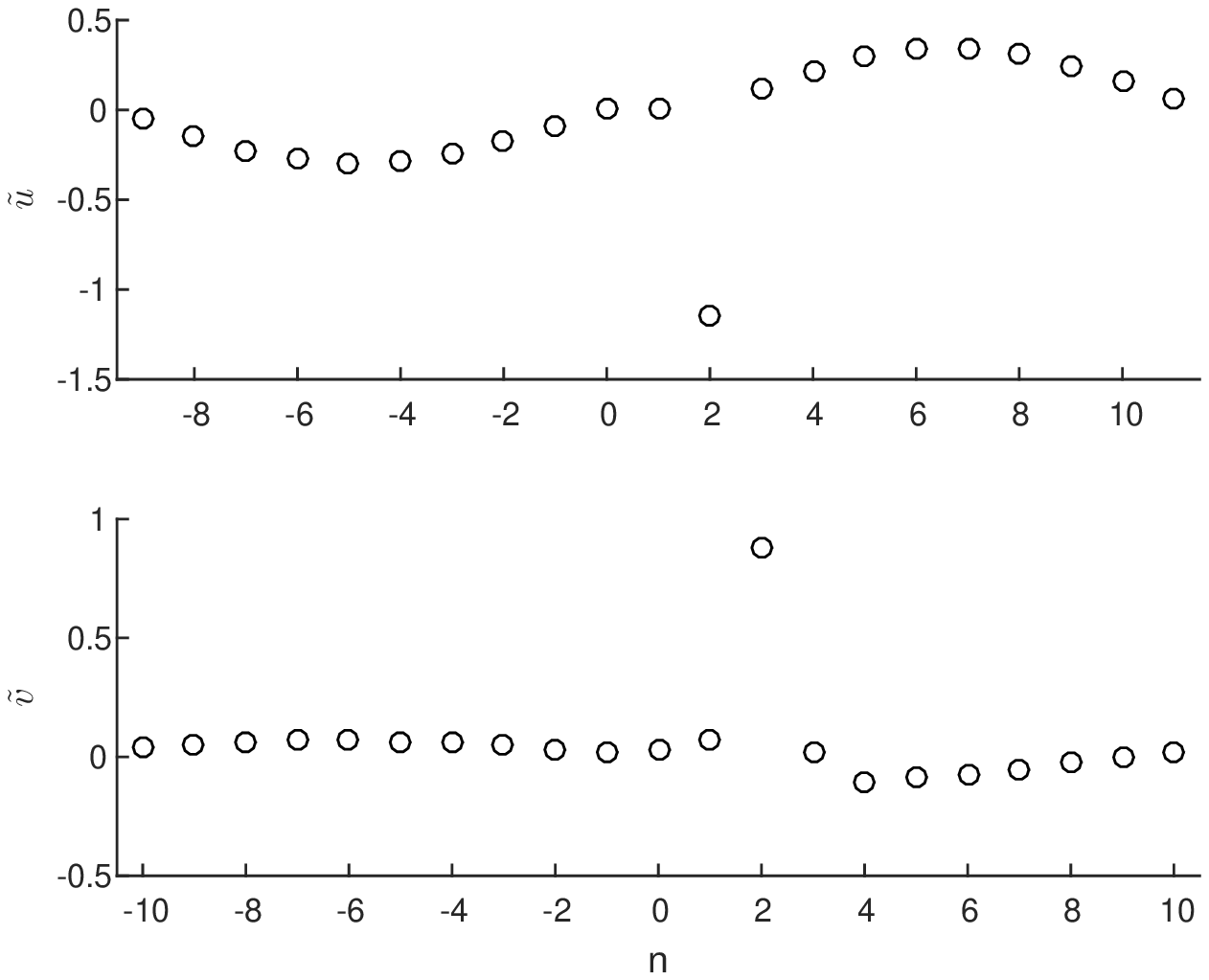}\\
\includegraphics[width=0.4\linewidth]{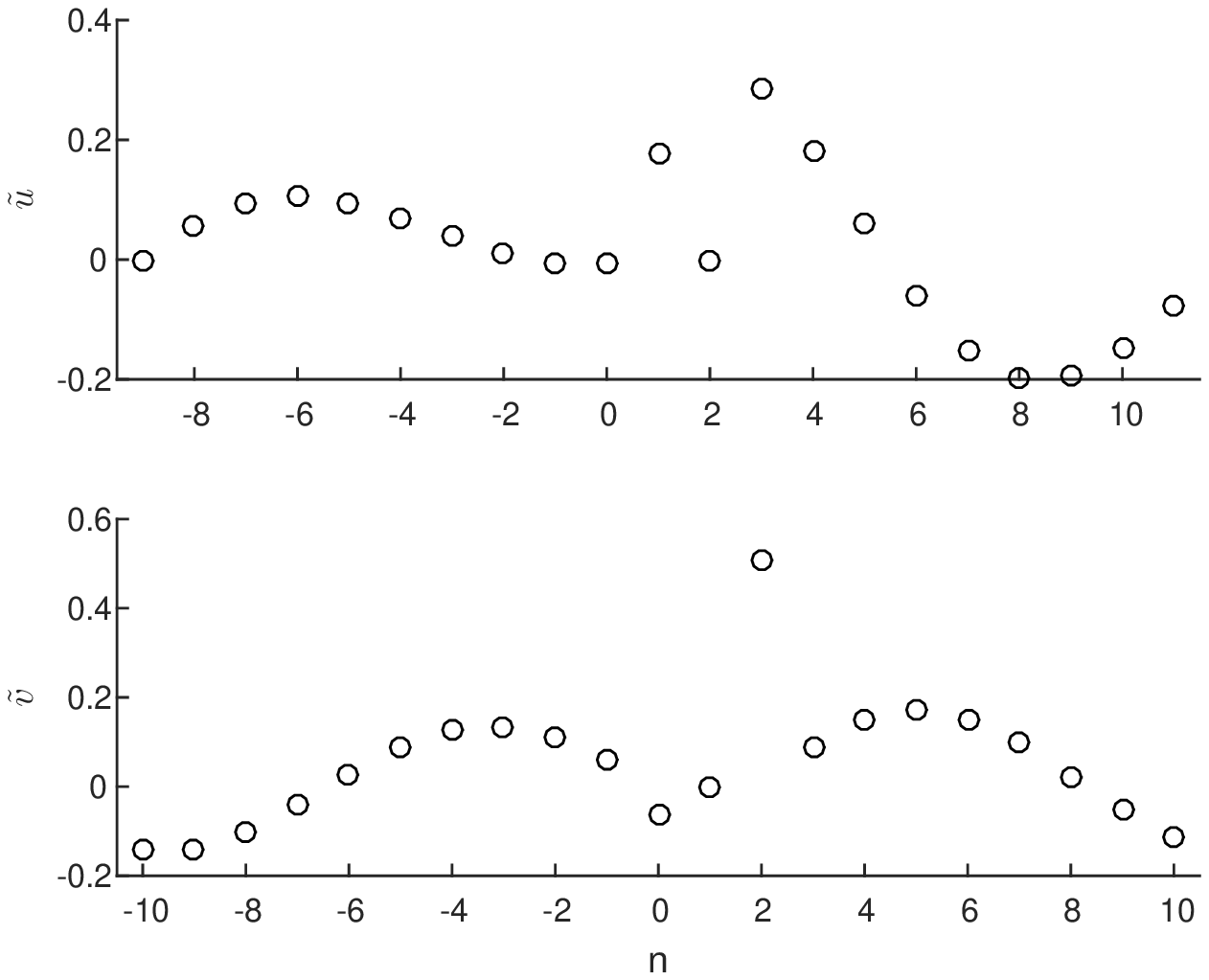}\includegraphics[width=0.4\linewidth]{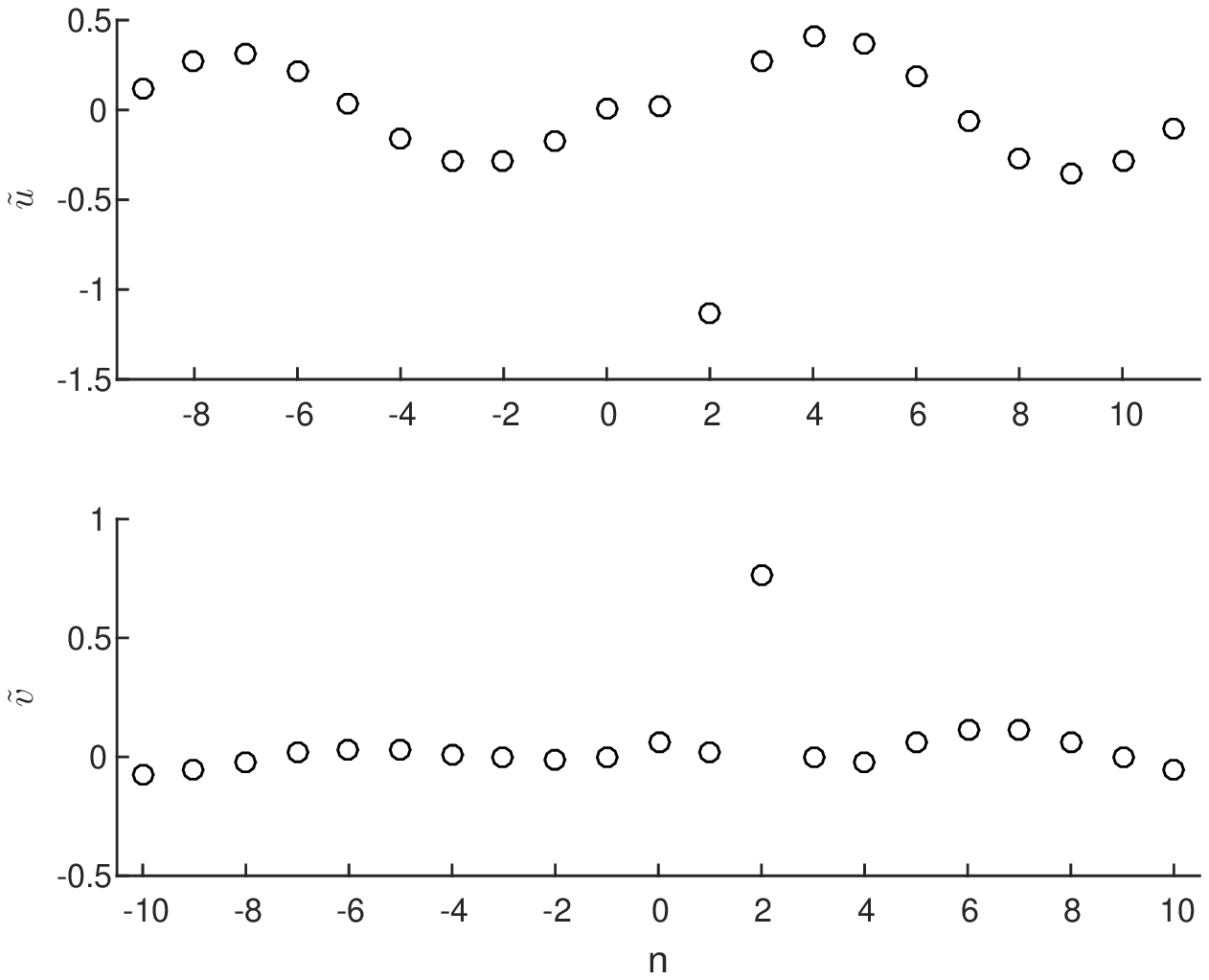}
\par\end{centering}

\protect\caption{\label{fig:18}Shapes of selected eigenvectors corresponding to Neimark-Saker
bifurcation for a 2-site MB with $N=20$ and sites at $0$ and $2$
in anti-phase corresponding to loss of stability stripes from the
top down in fig. \ref{fig:17} (left-to-right, top-to-bottom)}
\end{figure*}

Moreover, in the case of the conservative MBs the eigenvectors corresponding to the loss of stability also could be delocalized. In Fig. \ref{fig:17}
we demonstrate the stability diagram for the two-site anti-phase multi-breather.
This diagram demonstrates an interesting pattern of thin stability
strips, each boundary corresponding to the Neimark-Saker bifurcation with different
eigenvectors, as shown in Fig. \ref{fig:18}. This structure exhibits
strong dependence on the number of particles in the chain fragment, similar to the case
of the forced-damped DBs.

\section{\label{sec:VI}Concluding remarks}

The results presented above demonstrate, first of all, that one can analytically derive the exact solutions for the MBs and further explore their existence/stability patterns in the space of parameters with moderate computational efforts. The procedure also revealed certain complications, absent in the case of the single-site DBs. For instance, in order to study the phase differences between various localization sites, one should relax  the symmetry conditions and explore more generic families of the periodic localized solutions. Such undertaking would be a natural extension of current study.

Another interesting finding is strong dependence of the stability boundary for the Neimark-Saker bifurcation scenario, and lack of such dependence - for the pitchfork bifurcation. This study used periodic boundary conditions. In the real experiments, with free or fixed boundary conditions, one should expect significant dependence of the breather stability not only on the system size, but also on the proximity of the breather to the system boundary. This point also requires additional exploration.
\begin{acknowledgments}
The authors are very grateful to Israel Science Foundation (grant 838/13) for financial support.
\end{acknowledgments}

\bibliography{bibtex}

\end{document}